\documentclass[11pt]{article} 
\usepackage{hyperref}
\pdfoutput=1

\usepackage[T1]{fontenc}
\usepackage[utf8]{inputenc}
\usepackage{authblk}

\title{Interfacial Instabilities in Torsional Flows}
\author[1,3]{C.-Y. Lai}
\author[3]{Y.-T Sun}
\author[3]{C.-C. Chang}
\author[1]{Y.-Y. Chen}
\author[2]{P. E. Arratia}
\author[3]{J.-C. Tsai}

\affil[1]{Department of Physics, National Taiwan University, Taipei, Taiwan}
\affil[2]{Department of Mechanical Engineering, University of Pennsylvania, Philadelphia, Pennsylvania 19104, USA}
\affil[3]{Institute of Physics, Academia Sinica, Taipei, Taiwan}

\begin{document}
 \maketitle

We report on current findings on the morphology of an oil-water interface in a torsional flow produced by rotating the upper lid of a cylindrical tank. Here, the upper half of the tank is filled with oil and the lower half is filled with water. The interface morphology is investigated as a function of the Reynolds number (Re), based on the upper fluid (oil), and of aspect ratio (AR) between the cylinder's height and radius (AR=H/R). 

In the linked fluid dynamics \href{http://db.tt/5WrdcMgF}{video}, an upper fluid layer (silicone oil, 60cP) and a lower layer (water) are placed in a stationary cylinder (radius = 7cm, AR=H/R=1.4). To visualize the flow pattern, we add particles coated with silver (mean radius= 10 micrometers) which reflects the light when passing through a vertical laser sheet. As we spins up the upper lid to 130 rpm (Re=1111), the interface is quickly set into rotation and lifted up by the upward secondary flows in the oil. After a transient time, the lifted and curved interface is flattened by another downward flow developed near the central axis, and forms a very flat plateau which stays at a certain height. When we fixed the angular velocity, the shape of the interface stays stable. Thus, we can clearly classify different type of interface shape for a driven angular velocity and AR. 

However, when we increase the angular velocity beyond certain value, the interface suddenly breaks the axial symmetry, as the video shows next. In this case we use cooking oil (53cP) as the upper layer. These kind of symmetry breaking could also be observed in the rotating free surface confined in a stationary cylinder.\cite{2} The vedio shows symmetry breaking with three spikes developed along the edge of the plateau. On the tip of these spikes water bubbles break up from the interface and are continuously drawn up to the oil layer. As we change the parameter, we also found four, five and six spikes. All the configurations are rotating with a constant angular velocity lower than the rotating lid. Sometimes these nonaxisymmetric configurations are not stable. As shown in the vedio, we had observed seven spikes change gradually to three spikes, while the parameters (Re and AR) are fixed. The number of spikes depends very sensitively to the initial condition. All of these phenomena are found at AR ranges from 1 to 2 and Re from 970 to 2500.

Finally, we also show, at low Re(=2436) and high AR(=4.7), how the rising interface get squeezed and draws a long water thread up into the oil layer (silicone oil, 60cP). After the initial transient, the thread breaks into water bubbles, which could continuously be drawn from the interface as well. The entrained bubbles from the lower fluid to the upper one appears above a critical Re.  Interfacial tension plays an important role to determine the critical Re. The withdraw of fluid above an interface, entraining the lower fluid with the upper one exhibits similar behavior, but without rotation.\cite{3} These water bubbles, confined in the vortex ring, are useful  to visualize the flow pattern when the flow is steady, usually at low Re and low AR.

\end{document}